\documentclass[usenatbib]{mn2e}
\usepackage{epsfig,float,rotating,amsmath,flushend}

\def\5{\footnotesize V\normalsize}
\def\4{\footnotesize IV\normalsize}
\def\3{\footnotesize III\normalsize}
\def\2{\footnotesize II\normalsize}
\def\1{\footnotesize I\normalsize}

\def\kms{$\mbox{km s}^{-1}$}

\title[Sher\,25 pulsations]
{Sher\,25: pulsating but apparently alone}
\author[W. D. Taylor {\rm et al.}]{William D. Taylor$^{1}$, Christopher J. Evans$^1$, Sergio Sim\'{o}n-D\'{i}az$^{2,3}$, Hugues Sana$^{4}$, \newauthor 
Norbert Langer$^5$, Nathan Smith$^6$, Stephen J. Smartt$^7$ \vspace{0.2cm}\\
$^1$UK Astronomy Technology Centre, Royal Observatory Edinburgh, Blackford Hill, Edinburgh, EH9 3HJ, UK\\
$^2$Instituto de Astrof{\'i}sica de Canarias, 38200 La Laguna, Tenerife, Spain\\
$^3$Departamento de Astrof\'{õ}sica, Universidad de La Laguna, Avda. Astrof\'{õ}sico Francisco S\'{a}nchez s/n, E-38071 La Laguna, Tenerife, Spain\\
$^4$European Space Agency, Space Telescope Science Institute, 3700 San Martin Drive, Baltimore, MD 21218, USA\\
$^5$Argelander-Unstitut fur Astronomie der Universitat Bonn, Auf dem Hugel 71, 53121 Bonn, Germany\\
$^6$Steward Observatory, University of Arizona, 933 North Cherry Avenue, Tucson, AZ 85721, USA\\
$^7$Astrophysics Research Centre, School of Maths and Physics,Queen's University Belfast, Belfast BT7 1NN, UK
}

\begin{document}

\maketitle

\begin{abstract}
  \noindent The blue supergiant Sher\,25 is surrounded by an
  asymmetric, hourglass-shaped circumstellar nebula, which shows
  similarities to the triple-ring structure seen around SN1987A. From
  optical spectroscopy over six consecutive nights, we detect periodic
  radial velocity variations in the stellar spectrum of Sher\,25 with
  a peak-to-peak amplitude of $\sim$12\,\kms\,on a timescale of about 6\,days, confirming the tentative detection of
  similar variations by Hendry et al. From consideration of the amplitude and timescale of the signal, coupled with observed line 
  profile variations, we propose that the physical origin
  of these variations is related to pulsations in the stellar
  atmosphere, rejecting the previous hypothesis of a massive,
  short-period binary companion. The radial velocities of two other
  blue supergiants with similar bipolar nebulae, SBW1 and HD\,168625,
  were also monitored over the course of six nights, but these did not
  display any significant radial velocity variations.
\end{abstract}

\section{Introduction}

Many of the known Luminous Blue Variable (LBV) stars have associated
emission nebulae \citep[e.g.][]{hd94,nota95}. These are thought to
have originated from the large amounts of material thrown-off by the
parent star during violent episodes of mass-loss. The most famous
example is the Homunculus formed by the 19$^{\rm th}$ century eruption
of $\eta$ Carinae which, in common with several other LBV nebulae, has a
highly bipolar structure.

\begin{figure*}
  \begin{center}
       \includegraphics[width=17.7cm]{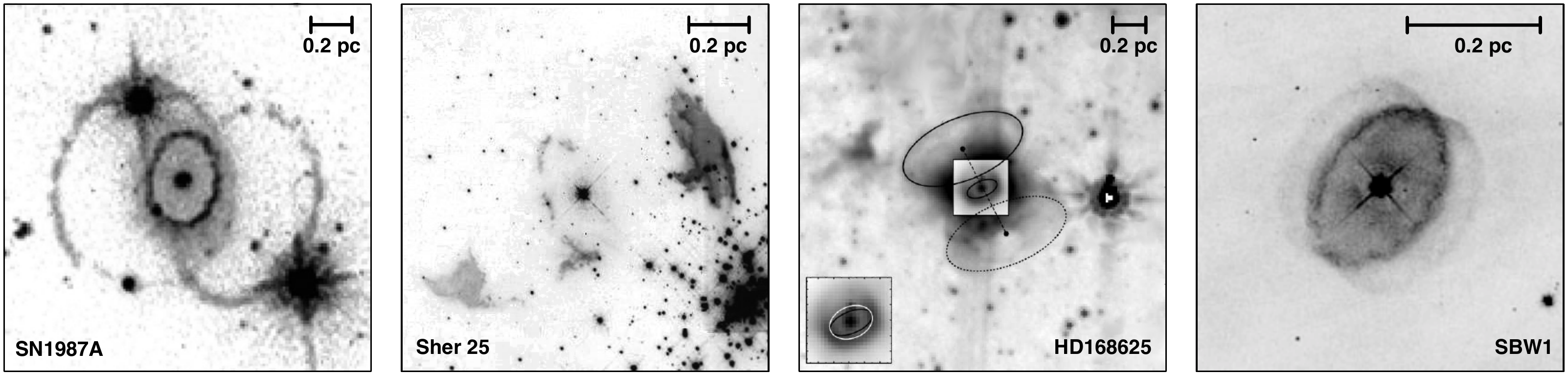} 
  \end{center}
  \caption{Images of the famous triple-ring nebula of SN1987A and the three LBV candidates observed in this study. SBW1 also has three well-defined rings making it the closest analogue of SN1987A, while Sher\,25 has large lobes of material aligned along the axis of the central ring. The rings of HD\,168625 are faint, so have been highlighted in this $Spitzer$ image from \citet{smith_HD168625}. Other image credits: SN1987A, $HST$, \citet{sn1987a_image}; Sher\,25, $HST$, \citet{brandner}; and SBW1, $HST$, \citet{sbw1_2013}. }\label{images}
\end{figure*}

Observations with the {\em Hubble Space Telescope} revealed that the
progenitor of SN1987A, Sk\,$-$69$^\circ$202 \citep[classified as
B0.7-3~I by][]{nolan_87a}, was surrounded by a bipolar nebula that
pre-dated the supernova explosion by about 20,000 years \citep{sn1987_rings}. This comprises
a well-defined equatorial ring and two larger rings that are roughly plane-parallel with the inner ring but offset along the system's polar axis (see Figure~\ref{images}).  Indeed, the
similarity of the nebula around SN1987A to that seen around the
candidate LBV HD\,168625, led \citet{smith_HD168625} to suggest that
Sk\,$-$69$^\circ$202 may have been a low-luminosity quiescent LBV, or
that it had at least undergone an LBV-like eruption in its recent past.

The LBV candidate Sher\,25 is a B-type supergiant slightly to the
northeast of the young massive cluster NGC\,3603 \citep{sher}; classified as B1.5 Iab \citep{moffat_sher}. 
It too has an associated nebula, consisting of a well-defined central ring
with two asymmetric lobes of material on either side \citep[][see
Figure~\ref{images}]{brandner}. The similarities in their
spectral types and associated nebulae have led to comparisons of
Sher\,25 with SN1987A \citep[e.g.][]{brandner2,smartt_2002}.

There is no clear consensus as to how the ring structures surrounding these stars were formed.  One of the most successful models is that of a binary merger between a 15\,$M_{\odot}$ star and a 5\,$M_{\odot}$ companion \citep{MP_2009}. Alternatively, \citet{chita} argued that a single star could generate this complex structure through the interaction of its stellar wind with material deposited during earlier stages of the star's evolution. Both of these models require the star to have passed through a red supergiant (RSG) phase, whereas a more recent theory by \citet{sbw1_2013} is independent of previous evolutionary stages. In this model the equatorial ring has been formed in a previous unexplained mass-loss event, while the subsequent photoevaporation of material from this ring interacting with the stellar wind forms the polar rings. Whether the progenitor of SN1987A and Sher\,25 have passed through a RSG phase cannot currently be determined conclusively.

In the course of investigating the abundances of Sher\,25 and its
nebula, \citet{hendry} detected radial velocity (RV) shifts with peak-to-peak amplitude of
$\sim$12\,km\,s$^{-1}$ from the five stellar observations available. The
RV shifts were consistent with either a three or six day period of a
binary companion. \citeauthor{hendry} noted that the implied secondary
in such a scenario would have to be more massive than Sher\,25 itself,
and yet we see no evidence for such a companion - for instance, \citet{naze} found no significant X-ray
emission from Sher\,25, which might be expected if a massive short-period companion were present. \citeauthor{hendry} speculated that
a binary with a centre-of-mass within the radius of Sher\,25 might
account for the observations (i.e. a slow merger), but also noted that
the RV shifts may have arisen from pulsations or some other
process in the stellar envelope.  Indeed, their reported
RV shifts were roughly the same magnitude as their measurement uncertainties,
although no equivalent shift was seen in the interstellar Ca~\2
absorption suggesting that the signature was real. 

If Sher\,25 has a short-period binary companion it could have important consequences for our understanding of the formation of its nebula. Likewise, the presence and type of pulsations within the stellar atmosphere could potentially indicate different evolutionary histories \citep{saio_2013}. Here we
report on a comprehensive set of follow-up observations to investigate
the nature of Sher\,25 in more depth. 

We also report on observations of two other LBV candidates with
similarities to SN1987A and Sher\,25, namely HD\,168625
\citep[e.g.][]{smith_HD168625} and SBW2007 1, commonly known as SBW1 \citep{darkness}.  Both are also
surrounded by triple-ring structures (see Figure~\ref{images}).
Classified as B6\,Iap \citep{nolan_2000}, HD\,168625 has a luminosity
comparable to that estimated for Sk\,$-$69$^\circ$202, while \citet{sbw1_2013} argue that the nebula of SBW1 is the closest known analog to that of SN1987A.  

\section{Observations}
The three stars were observed over six consecutive nights (spanning
March 19-24$^{\textrm{th}}$ 2009) using FEROS on the Max Planck Gesellschaft
(MPG)/European Southern Observatory (ESO) 2.2-m telescope at La Silla.
FEROS is a fixed-configuration instrument, with a wide wavelength
coverage of $\lambda\lambda$3600--9200\,\AA\ at a spectral resolving
power of 48\,000. Sher25 and SBW1 are sufficiently faint ($V$\,$=$\,12.3 and 12.7,
respectively) that pairs of back-to-back exposures of 2400\,s were
obtained. HD\,168625 ($V$\,$=$\,8.4) was observed in a series of
shorter back-to-back exposures (of 240 to 600\,s). A full list of the
observations is given in Table A1.

The data were reduced using the ESO Data Reduction Software pipeline
for FEROS.  Further steps can be taken beyond the pipeline reductions
to improve the merging of the spectral orders, but this was not
necessary for our purposes.  Also, our targets are sufficiently bright
that the spectrum from the FEROS sky fibre was not subtracted; indeed,
subtraction of the sky fibre actually degraded our data given the
additional source of noise.

\begin{figure*}
\centering
\includegraphics[width=18.0cm, height=9.5cm]{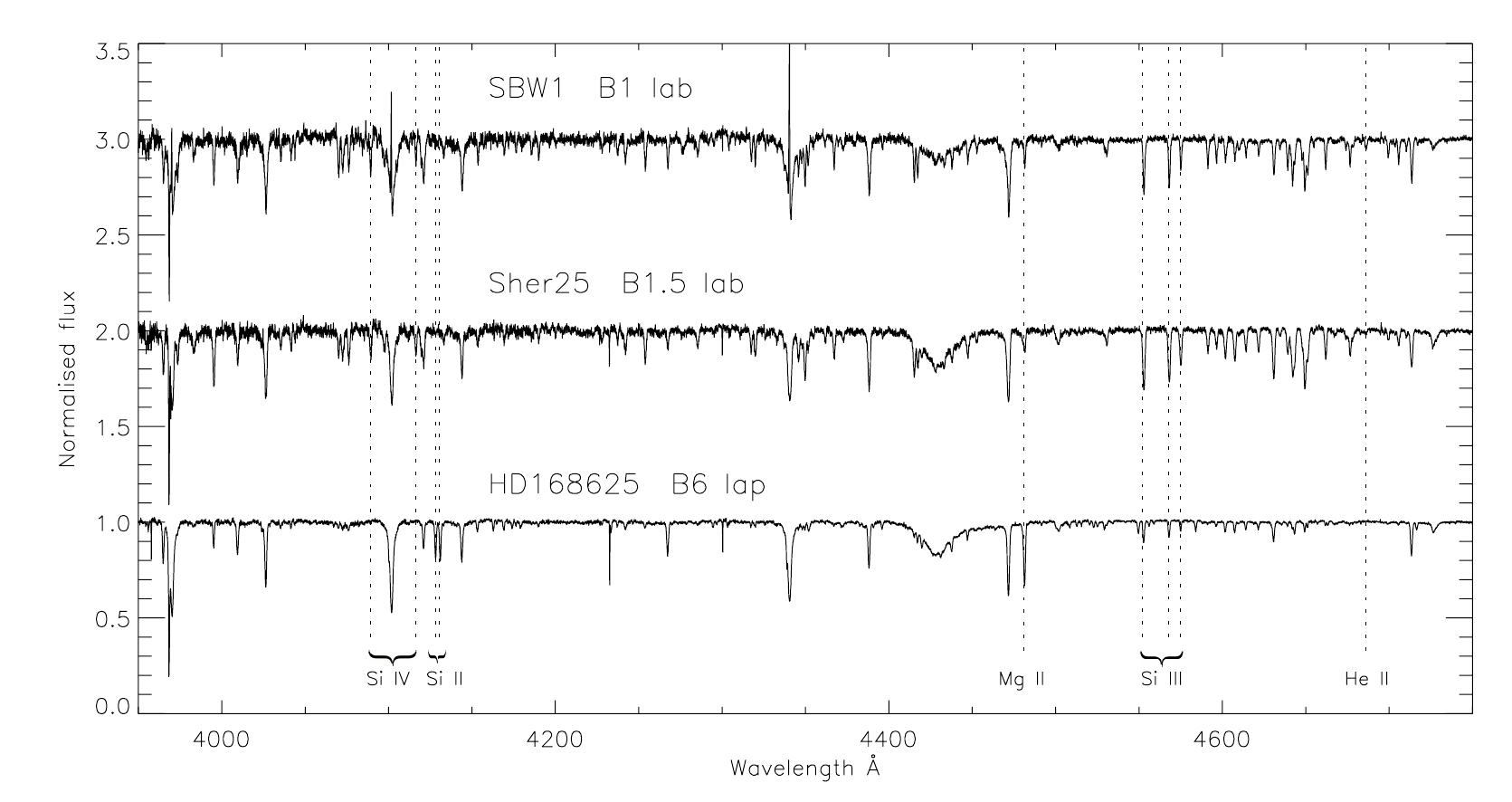}
\caption[]{
The blue-visual region of the median spectra of the three B-type supergiants observed with FEROS. For clarity the spectra have been smoothed using a
5-pixel, boxcar rolling average. The dashed lines indicate some of the most temperature sensitive lines for B-type supergiants. 
}\label{spectra_triptych}
\end{figure*}

Cosmic rays were removed from the individual spectra via comparisons
with a median-averaged spectrum of all the observations of each star.
The flux ratio of each spectrum to the median was calculated and a
boxcar 5$\sigma$-clip was applied over 100 wavelength bins to identify
and remove likely cosmics. Note that the median-averaging of the
spectra means that significant RV shifts in spectral lines between
frames could lead to regions being flagged as `suspect' pixels.  Thus,
only suspect pixels which were 5$\sigma$ {\em greater} than the local
continuum were removed, ensuring that any absorption features were
preserved intact.

The cleaned spectra were rectified by division of a low-order
polynomial fit to the continuum.  Back-to-back exposures of Sher\,25
and SBW1 were coadded to increase the signal to noise ratio (S/N), while all of the exposures
of HD\,168625 from a given night were coadded together as they were
obtained in quick succession. The typical S/N achieved in the final
spectra (measured around $\lambda4560$\,\AA, in the middle of the Silicon triplet) was $\sim$50 for Sher\,25 and SBW1, and $\sim$120 for
HD\,168625; example spectra of are shown in Figure~\ref{spectra_triptych}. 
The ratio of the Si\,IV to Si\,III lines suggest that SBW1 is marginally hotter than Sher\,25, which is also suggested by the presence of He II $\lambda 4686$ in SBW1's spectra. We therefore propose that SBW1 is reclassified as B1 Iab, while Sher\,25 retains its original classification of B1.5 Iab from \citep{moffat_sher}, noting the slight disagreement with \citet{melena_3603}, who classified Sher\,25 as B1 Iab.

\subsection{Radial velocity measurements }
Estimated RVs for each spectrum were obtained from two different
methods: profile fitting of sets of absorption lines and
cross-correlation (of sub-sections) of the spectra from different
epochs. In estimating the RVs, different regions were employed from the
broad coverage of the FEROS observations to ensure that any measured
signal was derived from different spectral orders (reducing the
possibility that an observed signal could be instrumental in origin). Also, lines from a number of different atomic species were used in the analysis, a full list of which is given in Table A1.

Prior to profile fits, a localised correction of the continuum
rectification was undertaken for each selected line using small
regions on either side of the absorption profile. The selected lines were
fit in the median-averaged spectrum initially.  This defined a median
width and depth of each line, which could then be held constant for
the fits to the individual (noisier) spectra, thus improving the
precision of the resulting RV estimates. For a number of lines the S/N 
was low and so not all lines were included in the final analysis, as indicated in Table A1 at the end of the paper.

As an independent check on the line-fitting analysis, measurements were also made using cross-correlation techniques. 
Three spectral regions were selected, each of which contained
well-resolved absorption lines available for inter-comparison:
$\lambda\lambda$4500-4700, $\lambda\lambda$5650-5750 and $\lambda\lambda$8580-8680\,\AA.

\begin{figure}
\centering
\includegraphics[width=8.5cm,height=4.7cm]{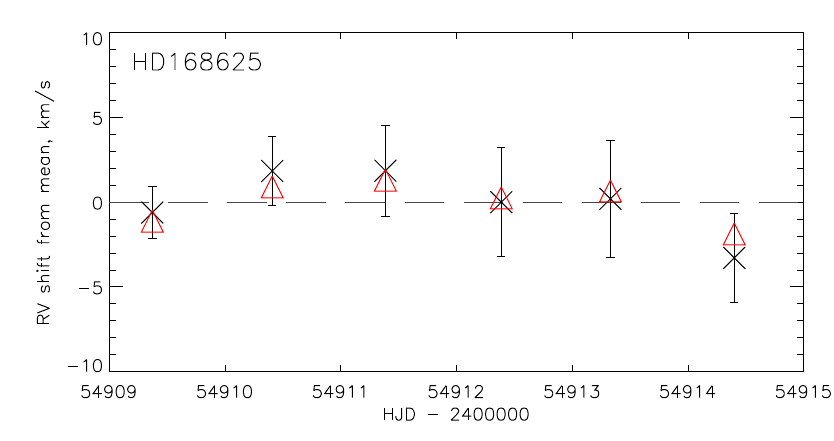}

\vspace{0.65cm}

\includegraphics[width=8.5cm,height=4.7cm]{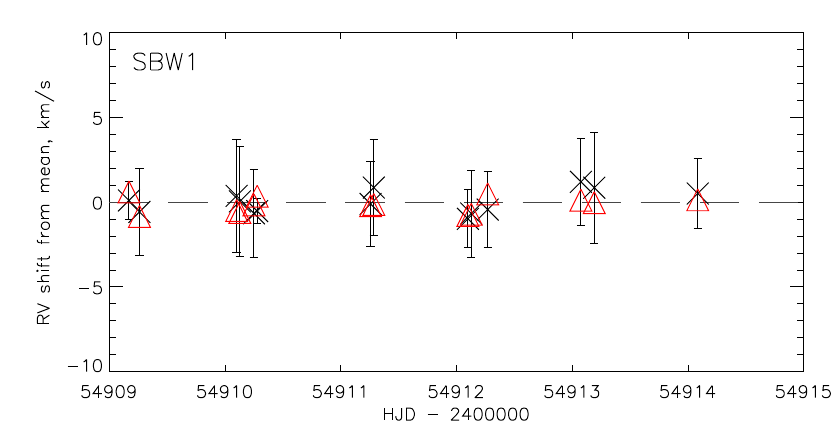}

\caption[]{Radial velocity measurements for HD\,168625 and SBW1. Black crosses:
  mean RV shifts estimated from Gaussian fits to spectral lines (with
  the 1$\sigma$ standard deviations indicated by the error bars).
  Red triangles: mean shifts estimated from cross-correlation
  measurements of the selected regions of the data. 
\label{other_shifts}}
\end{figure}

\begin{figure*}
\centering

\includegraphics[width=14.0cm]{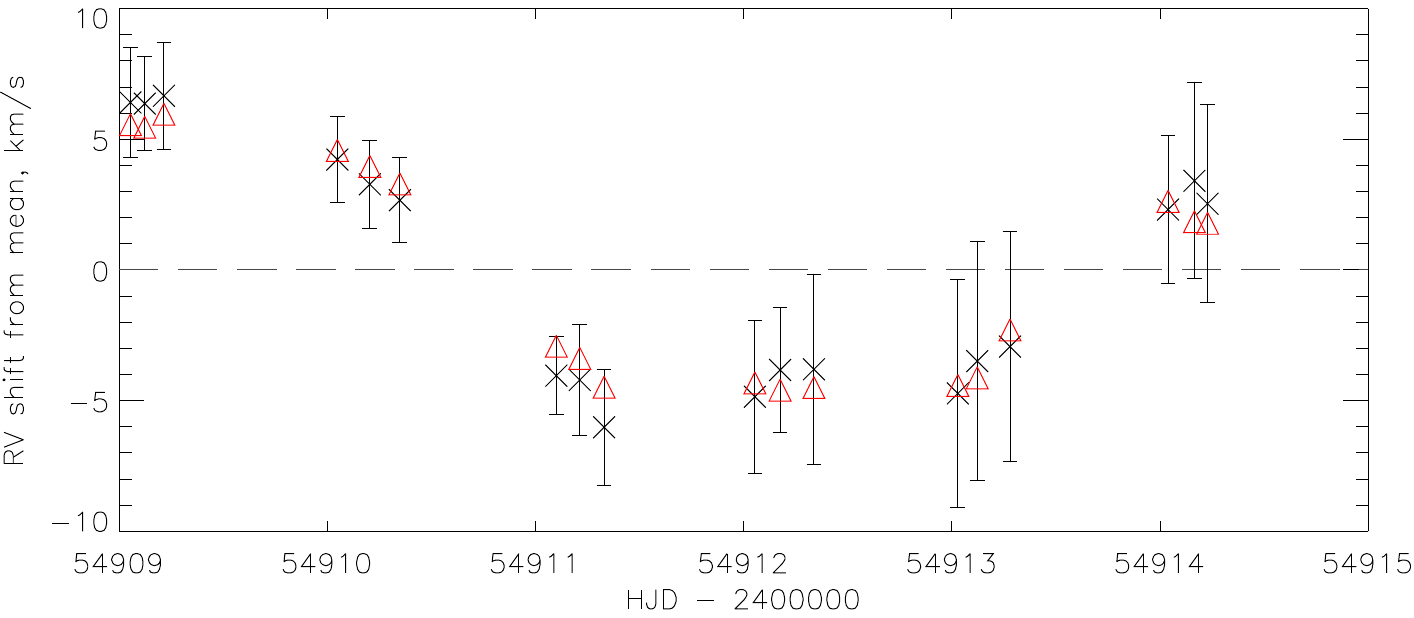}

\caption[]{Radial velocity measurements for Sher\,25. The
  symbols have the same meaning as those used in
  Figure~\ref{other_shifts}.  \label{sher25_shifts}}
\end{figure*}

The relative RV measurements for HD\,168625 and SBW1 are shown in Figure~\ref{other_shifts},
with those for Sher\,25 in Figure~\ref{sher25_shifts}; the full set of measurements is recorded 
in Table A1.  A clear RV variation is present in the Sher\,25 results from both the profile fits and from the
cross-correlation analysis. The magnitude of the RV variations is 
consistent with that reported by \citet{hendry}, but due to the delay since the previous observations, no attempt was 
made to include them in the period searches. 

Qualitative inspection of the results for HD\,168625 suggest a weak
trend with time (upper panel of Figure~\ref{other_shifts}), but any
deviation is within the associated uncertainties so we are unable to
draw firm conclusions from the current data.  We find no evidence for
significant short-term variations in SBW1 (lower panel of
Figure~\ref{other_shifts}).

\section{Interpretation of RV shifts}
In the following sections we consider possible scenarios
to explain the apparent RV variations detected in Sher\,25's
spectra. We begin by investigating binary companions as this was the initial motivation for this study. However, we note that the amplitude of the
detected variations is modest, particularly in light of results for
B-type supergiants in Westerlund\,1 from \citet{ritchie}, who found
$\Delta$RV\,$=$\,15--25\,\kms\ for some of their targets for
which they were unable to distinguish between orbital and photospheric
effects.

\begin{table}
\caption[]{Measured RV shifts for previous FEROS observations of Sher\,25. The mean shifts use the same zero-point as the values listed in Table A1.
\label{old_data} } 
\begin{minipage}{8.5cm}

\begin{center} 

\begin{tabular}{lll}
\hline\hline 
Date                 & Mean RV shift  & Error \\
(HJD - 240000) & (\kms)             & (\kms) \\
\hline

51327.000     &   \,\,3.20  &    6.00 \footnote{For this epoch the measured value and error are from \citet{hendry} adjusted to the same zero-point as the other epochs. }  \\ 
53191.092     &  -3.81  &    3.63 \\   
53193.054     &   \,\,2.36  &    3.81 \\
53369.852     &   \,\,3.17  &    3.24 \\
53370.852     &  -7.62  &    3.06 \\
53371.828     &  -8.15  &    2.58 \\

\hline
\end{tabular} 
\end{center} 
\vspace{-0.5cm}
\end{minipage}

\end{table} 

\subsection{Credible companions?}
An estimate of the periodic timescale for the RV signal was found from a basic Fourier analysis to be $\sim$6 days. Initial inspection of the data might suggest that the observed RV shift could represent the peak of a longer, larger oscillation: however, testing showed that the first night's observations constrain the fit to this apparent 6-day timescale. While caution should be exercised when assuming the observed variation is truly periodic, in the analysis that follows we consider the impact of these RV signals as if they were a periodic pattern - an assumption that is tentatively supported by the similarities to previous observations (see Table \ref{old_data}).

The 6-day period was used as the initial input to the Li\`ege Orbital Solution Package \citep[see][]{sana06stars}, to fit an orbital solution to the data assuming that Sher\,25 is an elliptical SB1 system \citep[using the well-established techniques of][]{wolfe_orbits}. The salient parameters from this fit being a putative period of 6.1 $\pm$ 0.7\,days with an amplitude of 6.12 $\pm$ 0.28\,km\,s$^{-1}$. Note that it is not possible to constrain the period more tightly as the timebase of the observations is comparable to the apparent period.

From comparisons of their model atmosphere results with the
evolutionary tracks of \citet{meynet}, \citeauthor{hendry} estimated
the present-day mass of Sher\,25 to be 40\,$\pm$\,5\,$M_{\odot}$with a radius of 60\,$\pm$\,15\,$R_{\odot}$. Adopting this mass estimate for the
primary, assuming an inclination of 65$^\circ$ from the measured tilt of the central ring \citep{brandner} and using the results from the orbital fit, results in an estimated secondary mass of $\approx$0.66\,$\pm$\,0.2\,$M_{\odot}$. This contrasts with the suggested high-mass companion of \citet{hendry} as we have measured a longer period.

Following Keplerian arguments the orbital separation of the two hypothetical
components is calculated to be $\sim$40\,$\pm$\,14\,$R_{\odot}$;
i.e. a companion would be very close to, or even within, the extended
envelope of the primary.  Similar calculations with the primary mass varied by up to $\pm$50\% (i.e.\,allowing for different evolutionary scenarios where Sher\,25 may have passed through a RSG phase), also suggest
a low-mass companion at a separation comparable to the estimated
radius of the primary. It therefore seems unlikely that the observed RV signal is the result of a binary companion.

\subsubsection{Other possible binary companions}
\begin{figure}
  \begin{center}
  \vspace{0.03cm}
    \includegraphics[width=8.4cm,height=5.8cm]{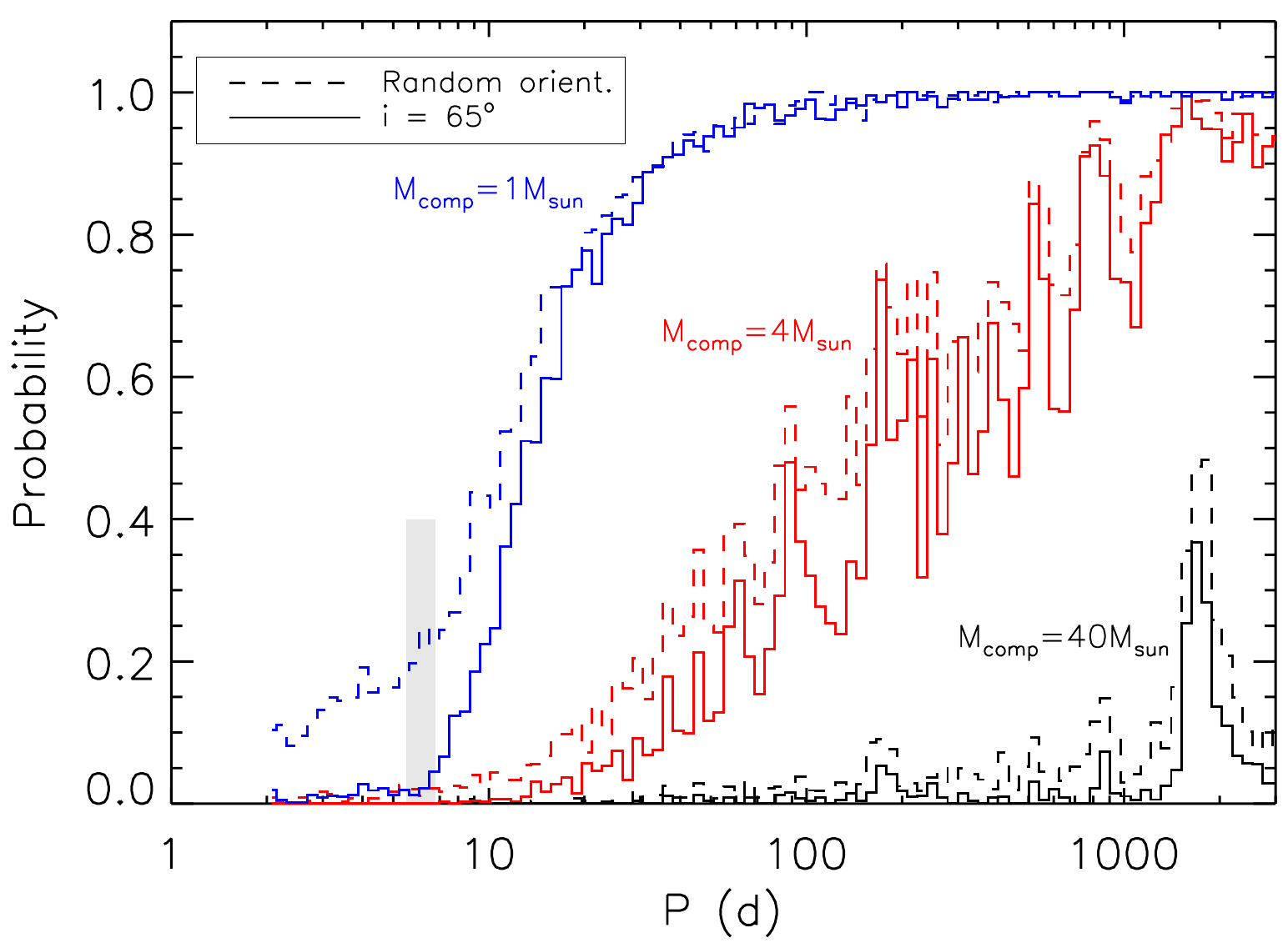}
  \end{center}
  \caption[]{
  Probability that a possible companion around Sher\,25 would generate a RV signal smaller than the detection threshold for these observations. The shaded band indicates the predicted 6.1\,day period and its associated error.
\label{completeness}}
\end{figure}

It is interesting to explore the orbital properties and companion mass that would be compatible with the RV signal detected for Sher 25. To do this we used the methods of Sana et al. (2009, 2011) to simulate the observable RV shifts that would be expected for various binary systems. Three different mass-ratios were modelled, each with a 40\,M$_{\odot}$ primary and either a 40\,M$_{\odot}$, 4\,M$_{\odot}$ or 1\,M$_{\odot}$ secondary. A range of orbital eccentricities consistent with those of \citet{sana_science} were considered. Predictions can then be made about the probability of detecting the expected RV shifts assuming the typical error of our measurements and the time-sampling of the observations. To further increase this time-sampling, the previous FEROS observations from \citet{hendry} were re-analysed using the methods described above (for completeness, the measured RV shifts of these additional observations are detailed in Table~\ref{old_data}). Figure \ref{completeness} shows the probability results for the three different mass-ratios and also the impact of assuming an inclination for the system of 65$^\circ$, as measured by \citet{brandner}. The probabilities represent confidence levels of 4.7$\sigma$, which equate to velocity shifts of 15\,\kms. 

Figure \ref{completeness} confirms that the properties of the RV signal detected (both RV amplitude and timescale) cannot be reproduced by a physically realistic companion. It is possible that one or even several companions exist beyond our detection threshold, i.e. at orbital separations large enough to produce a negligible signal in our data. Searching for such companions would require either long-term spectroscopic monitoring or very high-angular resolution techniques (Sana et al 2014, ApJS, subm.). The impact of such putative companions on the evolution of Sher\,25 would then depend on the companion's mass, eccentricity and orbital separation.

\vspace{-0.1cm}
\subsection{Wind variability?}
\begin{figure}
  \begin{center}
    \includegraphics[width=8.5cm,height=5.4cm]{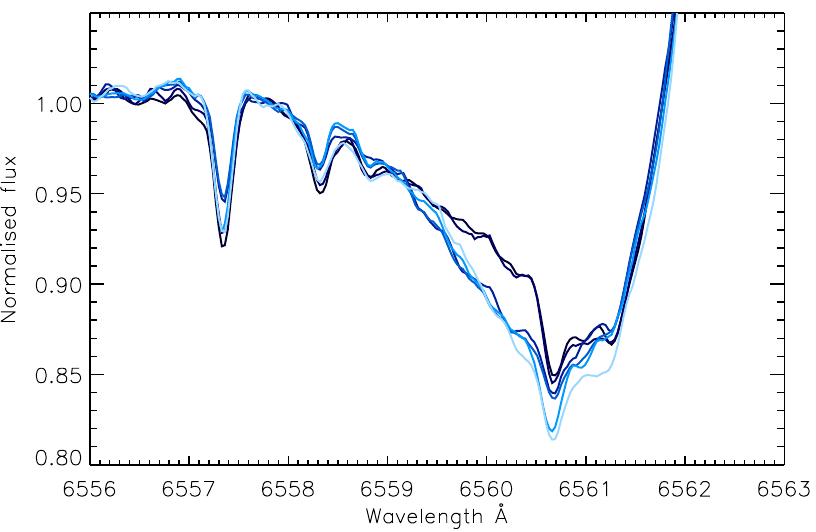}
    
    \vspace{0.3cm}
    \includegraphics[width=8.5cm,height=5.4cm]{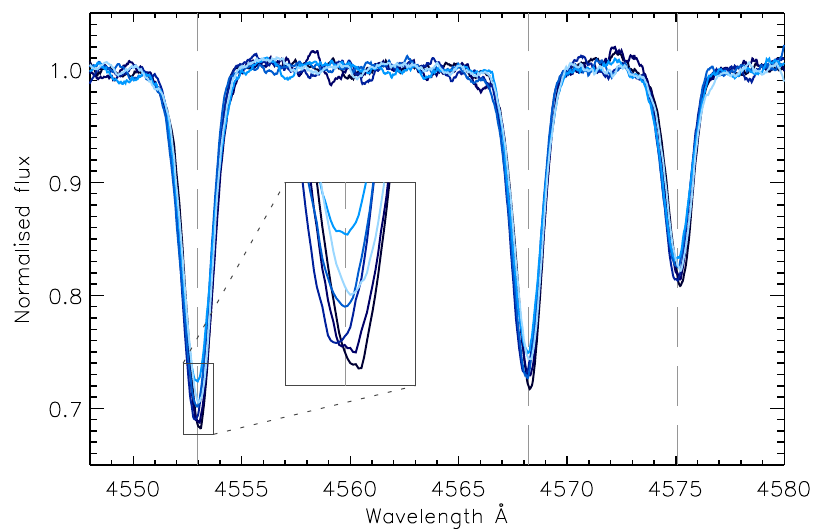}

  \end{center}
  \caption[]{ The upper panel shows the variation in the wind profile of the H$\alpha$ 
line for Sher\,25, while the lower panel shows the subtle changes in the profile shapes of the
Si~{\scriptsize III} triplet. The colour indicates the observation's date, going from black to cyan (i.e. dark to light) with increasing time. For clarity, the data from each night have been coadded and smoothed using a 5-pixel boxcar average.  \label{halpha}}
\vspace{-0.3cm}
\end{figure}

\citet{hendry} commented on variations in Sher\,25's
H$\alpha$ P-Cygni wind profile and similar variations are present
in the new FEROS data (see the upper panel of Figure~\ref{halpha}). This behaviour could
be caused by inhomogeneous large-scale structure within the wind,
leading to variations in the absorption profile \citep[e.g.,][]{markova}. The H$\alpha$ profiles do not display periodic behaviour within our
observations, and although the similarity with the previous data might suggest some longer period repeatability, it is clearly with a different timescale to the observed RV shifts. Wind variability is therefore unlikely to be the origin of Sher\,25's apparent RV signal. 

\subsection{Possible pulsations?}\label{checks}
A number of Sher\,25's metal lines appear to show small-scale variations. Such behaviour is seen in many OB-supergiants and has long been attributed to pulsations within the stellar envelope \citep{Lucy}. The lower panel of Figure~\ref{halpha} shows the variation in depth seen for the Si~\3 triplet of Sher\,25. There is a less obvious trend with time than in the H$\alpha$ profile --
the deepest profiles are from the first night, with the lines appearing
to get shallower until the final night, when the pattern reverses,
hinting at a more periodic behaviour.

\citet{balona} proposed that the temporal variation of the first three moments of a line-profile could be used to investigate pulsations in stars. Specifically,
the first moment provides a measurement of the centroid velocity, and
the third moment (`skewness') gives a measurement of any asymmetry in
the lines.  Following this strategy, one can better separate real RV variations, e.g. due to binarity, from those resulting from pulsations,
i.e. when the line centre is `wobbling' due to changes in the skewness and,
consequently, the measured velocity of the line also appears to move.

\begin{figure}
\centering
\includegraphics{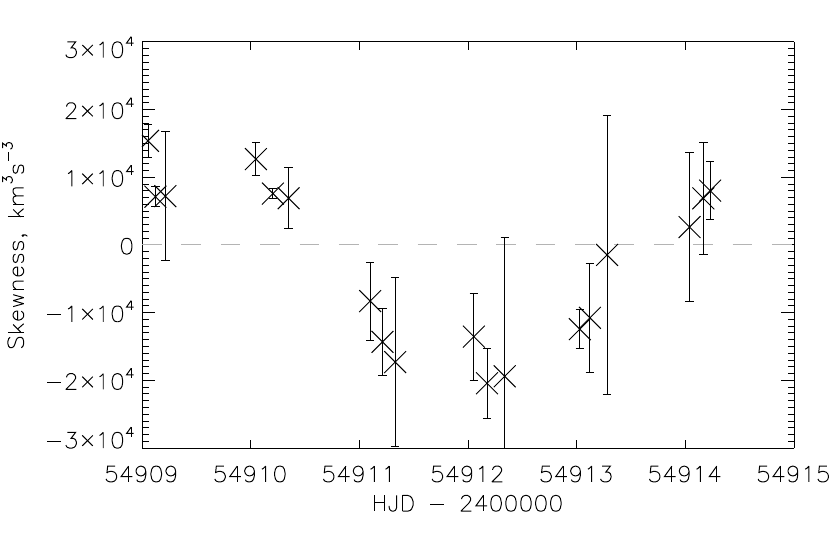}
\caption[]{
Skewness results for Sher\,25. The data show the mean skewness for each observation calculated from the Si~{\scriptsize III} triplet. The errors represent the standard deviation on the measurements of the three lines. 

\label{skewness}}
\end{figure}

Through the {\sc famias}\footnote{{\sc famias} was developed in the framework of the FP6 European Coordination Action HELAS (http://www.helas-eu.org/)} package \citep{zima_FAMIAS}, which employs the moment analysis techniques of \citet{briquet}, we computed the third moment of all the Si~{\scriptsize III} triplet lines (lower panel of Figure \ref{halpha}). The results are shown in Figure~\ref{skewness} and clearly suggest that there is periodic variation in the skewness of the lines, with a period similar to the measured RV variations. The order of magnitude and the time frame of these variations are consistent with those observed for other B-type supergiants by \citet{sergio_lpvs}, although the pattern is noticeably more periodic than the stars of described therein. It is possible that the relatively long exposures used for Sher\,25 have smoothed out possible smaller-scale variations. Our hypothesis is therefore that the observed RV variations in Sher\,25 are the result of line profile variations arising from stellar pulsations.

Similar skewness tests were performed for SBW1 and HD\,168625, with neither showing a variation in the skewness results compared to that for Sher\,25. This is consistent with the lack of detected RV variation in these two stars. Of course, this cannot rule out the presence of any undetected longer period variations for these stars.

To assess the type of pulsation seen in Sher\,25, V-band photometric data were obtained from ASAS\footnote{The `All Sky Automated Survey' provides constant photometric monitoring for all objects with V$<$14\,mag \citep{ASAS}.} to look for brightness variations that would imply radial pulsations. The ASAS data revealed no significant variation: indeed the standard deviation of the measurements was $<$0.03\,mag, which is comparable with the anticipated error on the ASAS measurements. Searches revealed no periodic signatures consistent with the predicted 6.1 $\pm$ 0.7\,day period. As an additional check, the Baade-Wesselink method can be used to estimate the typical RV-shifts associated with brightness variations. For a star with the properties of Sher\,25 and assuming a 6 day period, a 0.03\,mag variation would be expected to induce a RV-shift of $<$2\,\kms. We can therefore conclude that the observed RV signal of Sher\,25 is not associated with radial pulsations.

\subsubsection{Macroturbulence}
The line profiles of many luminous B-type stars cannot be accurately described simply by rotational broadening, an additional broadening mechanism - so called `macroturbulence' - must be invoked. \citet{aerts_2009} showed that the observed profiles could be generated through the superposition of numerous low-amplitude non-radial gravity mode oscillations within the stellar envelope. They predicted that the magnitude of this contribution should be proportional to any variation in the profile skewness, a result that has since been verified observationally by \citet{sergio_lpvs}. 

We calculated the macroturbulent velocity ($\Theta_{RT}$) of all the stars using the techniques of \citet{sergio_2013}, which exploits a combined Fourier transform and goodness of fit technique to disentangle the broadening contributions from rotation and macroturbulence. The results of the analysis are shown in Table \ref{macro_velocity}. For Sher\,25 the magnitude of these broadening terms and the results of the skewness analysis above, are consistent with both the predictions of \citet{aerts_2009} and the measurements of \citet{sergio_lpvs}, suggesting that in this respect, Sher\,25 is similar to other pulsating B-type supergiants.

The $\Theta_{RT}$\ estimate for SBW1 is not that much lower than that for Sher\,25; therefore, based on the predictions of \citet{aerts_2009}, we would have expected to find some evidence of skewness variation for SBW1. It is possible that the time-sampling of the observations is insufficient to detect the variation or that the signal has been smoothed out by the relatively long exposures. Understanding the relationship between pulsations and line broadening is still in its infancy; these results suggest that the frequency of pulsations will also prove an interesting area to explore further.

\subsubsection{Evolutionary implications of pulsations}
Through modelling crossed-checked against observations, it is possible to define evolutionary phases when supergiants are more likely to undergo pulsations. \citet{pam_1999} defined the regions in which stars might experience high-order g-mode instabilities, and later \citet{saio_2006} identified a region of stars exhibiting Post-TAMS low-order g-mode oscillations. The results of the {\sc fastwind} analysis for Sher\,25 performed by \citet{hendry}, suggest that the star lies on the boundary of these two instability regions; it is therefore not surprising that the star exhibits pulsations.

\citet{saio_2013} have taken these ideas further to argue that the evolutionary history of a star could be unravelled by the type of pulsation it exhibits. They argue that if a blue supergiant displays evidence for radial pulsations and lies within a certain region of the Hertzsprung-Russell Diagram (HRD), then it must have passed through a previous RSG phase and is currently on a blue-loop travelling back across the HRD. Unfortunately it is not possible to use these techniques to better understand Sher\,25's history as it lies just outside Saio's identified region and the absence of any significant photometric variation suggests that it does not exhibit radial pulsations. For SBW1, however, the luminosity and temperature estimates of \citet{sbw1_2013} suggest that the star lies within Saio's identified region, and while in no way conclusive, the absence of any pulsations for SBW1 supports the argument that the star has not passed through an RSG phase \citep{sbw1_2013}.

\begin{table}
\caption[]{Rotational and macroturbulent velocities derived from a combined Fourier transform and goodness of fit technique.  $v_{eq}$ gives an estimate for the stars' equatorial velocities based on inclinations derived from the central ring of the nebulae (see text for references). The quoted errors reflect the 3\,$\sigma$ confidence region of the fit and all the quoted velocities are in \kms.  
\label{macro_velocity} } 

\begin{center} 

\begin{tabular}[width=10cm]{l l l l l}
\hline\hline 
Star        		& $v\sin{i}$   		& $\Theta_{RT}$	& $v_{eq}$		 \\
\hline

Sher\,25     	&  	53 (+11/-8)       & 72 (+13/-11)		&  	64 (+13/-10) 	  \\
SBW1        	&  	34 (+23/-15)     & 55 (+21/-17) 		&  	41 (+28/-18) 	  \\
HD168625  	& 	44 (+6/-8) 		& 44 (+16/-16)		&  	53 (+7/-10)  	  \\

\hline
\end{tabular} 
\end{center} 

\end{table}

\section{Discussion}
It is interesting to consider whether the results reported here can influence the debate on the origin of the nebulae that surround these three stars and SN1987A. 

While no binary companions were found in this study, the estimates for detection probabilities of other companions presented in Figure \ref{completeness} indicate that it is quite possible that a smaller companion with a longer period could have gone undetected. Although, it should be noted that the entire lifetime of a B1 supergiant is comparable to the pre main-sequence lifetime of a solar-mass star. Therefore, if a small companion does now exist, it must have started life with a higher mass and subsequently had mass stripped from it, a process that would have significantly affected the evolution of both stars. 

If at a later date any companion were discovered, our analysis would suggest that it will either be a distant one, which may have had little impact on Sher\,25's evolution, or a small nearby companion, which has undergone significant interaction with Sher\,25. Frustratingly, this therefore does little to influence the nebula-origin debate.

Our treatment of macroturbulence has led to a lower estimate of the rotational velocity of 58\,\kms \,compared to the previous estimate of 108\,\kms \,from \citet{smartt_2002} - again this is assuming the inclination from \citet{brandner}. The rotational velocity of these stars is of particular relevance, as it could help assess the likelihood of different models. As an example, \citet{smith_townsend} express concern that the binary merger model of \citet{MP_2009} has insufficient time to slow the star's rotation since the merger. This problem is exaggerated by the slower $v\sin{i}$ estimates reported here. 

Interestingly, \citet{heger98} predicted that a single 25 M$_{\odot}$ star would deposit material in an equatorial ring as it began a blue-loop back across the HRD following a RSG phase. Their models predicted that the star's rotation would be dramatically slowed as angular momentum would be lost to the material deposited in the ring. They estimated that the final rotational velocity of the star would be around 50\,\kms, which is intriguingly close to that found for the stars in this study. This type of mechanism, coupled with the polar rings of \citet{sbw1_2013} that are formed through photoevaporation of material from the equatorial ring, could provide a possible solution for forming these nebula.

A final consideration is whether the stars have passed through a RSG phase. This is necessary for the models of \citet{heger98}, but for SBW1 the surface abundances suggest this is not the case - a conclusion which is tentatively supported by the pulsational theories of \citet{saio_2013} discussed above. Interestingly \cite{georgy} have expanded on Saio's work to show that through different modelling techniques a star could be seen to have lower than expected surface abundances even after passing through a RSG phase. This opens again the question of SBW1's history and shows that this crucial debate surrounding the possible RSG phase of these stars appears to be far from solved.

\section{Conclusion}
Due to its similarities to the mysterious SN1987A, Sher25 is an
excellent example of a very interesting breed of stars. This work
strongly suggests that the star is pulsating in a regular periodic
way, which mimics the RV signal of a small binary companion. Such a
companion is believed to be unlikely because the 
variation in the skewness of the line profiles matches the period of
the RV shifts, and also because any binary system with these orbital
parameters would be relatively unphysical. This work therefore lays to rest the possibility of a massive short-period
companion raised by \citet{hendry}. 

Both SBW1 and HD168625 appear to show no evidence for either RV variations or pulsations on the timescales of a few days considered here. All three stars are found to exhibit macroturbulent line-broadening, which lowers some of the previous estimates for their rotational velocity. These new estimates of rotation should provide additional constraining factors in the simulations of those seeking to explain the origin of these extraordinary nebulae. 

\ \\
\noindent {\sc acknowledgements:} Based on observations at the
European Southern Observatory primarily from programme 082.D-0136, but with additional analysis of observations from 071.D-0180 and 074.D-0021. With thanks to Selma de Mink for useful discussions about possible (or in fact, not possible) binary scenarios. SS-D acknowledges funding by the Spanish Ministry of Economy and Competitiveness (MINECO) under the grants AYA2010-21697-C05-04, Consolider-Ingenio 2010 CSD2006-00070, and Severo Ochoa SEV-2011-0187, and by the Canary Islands Government under grant PID2010119. SJS thanks the ERC and EU's (FP7/2007-2013)/ERC Grant agreement n$^{\rm o}$ [291222].

\nocite{sana_binarityIII}
\nocite{sana_binarityII}

\bibliographystyle{mn2e}

\bibliography{taylor_sher25}

\begin{thebibliography}{41}
\expandafter\ifx\csname natexlab\endcsname\relax\def\natexlab#1{#1}\fi

\bibitem[{{Aerts} {et~al}\mbox{.}(2009){Aerts}, {Puls}, {Godart}, \&
  {Dupret}}]{aerts_2009}
{Aerts} C., {Puls} J., {Godart} M., {Dupret} M.-A., 2009, \aap, 508, 409

\bibitem[{{Balona}(1986)}]{balona}
{Balona} L.~A., 1986, \mnras, 219, 111

\bibitem[{{Brandner} {et~al}\mbox{.}(1997{\natexlab{a}}){Brandner}, {Chu},
  {Eisenhauer}, {Grebel}, \& {Points}}]{brandner2}
{Brandner} W., {Chu} Y.-H., {Eisenhauer} F., {Grebel} E.~K., {Points} S.~D.,
  1997{\natexlab{a}}, \apjl, 489, L153

\bibitem[{{Brandner} {et~al}\mbox{.}(1997{\natexlab{b}}){Brandner}, {Grebel},
  {Chu}, \& {Weis}}]{brandner}
{Brandner} W., {Grebel} E.~K., {Chu} Y.-H., {Weis} K., 1997{\natexlab{b}},
  \apjl, 475, L45

\bibitem[{{Briquet} \& {Aerts}(2003)}]{briquet}
{Briquet} M., {Aerts} C., 2003, \aap, 398, 687

\bibitem[{{Burrows} {et~al}\mbox{.}(1995{\natexlab{a}}){Burrows}, {Krist},
  {Hester}, {Sahai}, {Trauger}, {Stapelfeldt}, {Gallagher}, {Ballester},
  {Casertano}, {Clarke}, {Crisp}, {Evans}, {Griffiths}, {Hoessel}, {Holtzman},
  {Mould}, {Scowen}, {Watson}, \& {Westphal}}]{sn1987_rings}
{Burrows} C.~J. {et~al.}, 1995{\natexlab{a}}, \apj, 452, 680

\bibitem[{{Burrows} {et~al}\mbox{.}(1995{\natexlab{b}}){Burrows}, {Krist},
  {Hester}, {Sahai}, {Trauger}, {Stapelfeldt}, {Gallagher}, {Ballester},
  {Casertano}, {Clarke}, {Crisp}, {Evans}, {Griffiths}, {Hoessel}, {Holtzman},
  {Mould}, {Scowen}, {Watson}, \& {Westphal}}]{sn1987a_image}
{Burrows} C.~J. {et~al.}, 1995{\natexlab{b}}, \apj, 452, 680

\bibitem[{{Chita} {et~al}\mbox{.}(2008){Chita}, {Langer}, {van Marle},
  {Garc{\'{\i}}a-Segura}, \& {Heger}}]{chita}
{Chita} S.~M., {Langer} N., {van Marle} A.~J., {Garc{\'{\i}}a-Segura} G.,
  {Heger} A., 2008, Astronomy and Astrophysics, 488, L37

\bibitem[{{Georgy} {et~al}\mbox{.}(2013){Georgy}, {Saio}, \& {Meynet}}]{georgy}
{Georgy} C., {Saio} H., {Meynet} G., 2013, \mnras

\bibitem[{{Heger} \& {Langer}(1998)}]{heger98}
{Heger} A., {Langer} N., 1998, \aap, 334, 210

\bibitem[{{Hendry} {et~al}\mbox{.}(2008){Hendry}, {Smartt}, {Skillman},
  {Evans}, {Trundle}, {Lennon}, {Crowther}, \& {Hunter}}]{hendry}
{Hendry} M.~A., {Smartt} S.~J., {Skillman} E.~D., {Evans} C.~J., {Trundle} C.,
  {Lennon} D.~J., {Crowther} P.~A., {Hunter} I., 2008, \mnras, 388, 1127

\bibitem[{{Humphreys} \& {Davidson}(1994)}]{hd94}
{Humphreys} R.~M., {Davidson} K., 1994, \pasp, 106, 1025

\bibitem[{{Lucy}(1976)}]{Lucy}
{Lucy} L.~B., 1976, \apj, 206, 499

\bibitem[{{Markova} {et~al}\mbox{.}(2005){Markova}, {Puls}, {Scuderi}, \&
  {Markov}}]{markova}
{Markova} N., {Puls} J., {Scuderi} S., {Markov} H., 2005, \aap, 440, 1133

\bibitem[{{Melena} {et~al}\mbox{.}(2008){Melena}, {Massey}, {Morrell}, \&
  {Zangari}}]{melena_3603}
{Melena} N.~W., {Massey} P., {Morrell} N.~I., {Zangari} A.~M., 2008, \aj, 135,
  878

\bibitem[{{Meynet} {et~al}\mbox{.}(1994){Meynet}, {Maeder}, {Schaller},
  {Schaerer}, \& {Charbonnel}}]{meynet}
{Meynet} G., {Maeder} A., {Schaller} G., {Schaerer} D., {Charbonnel} C., 1994,
  \aaps, 103, 97

\bibitem[{{Moffat}(1983)}]{moffat_sher}
{Moffat} A.~F.~J., 1983, \aap, 124, 273

\bibitem[{{Morris} \& {Podsiadlowski}(2009)}]{MP_2009}
{Morris} T., {Podsiadlowski} P., 2009, \mnras, 399, 515

\bibitem[{{Naz{\'e}} {et~al}\mbox{.}(2012){Naz{\'e}}, {Rauw}, \&
  {Hutsem{\'e}kers}}]{naze}
{Naz{\'e}} Y., {Rauw} G., {Hutsem{\'e}kers} D., 2012, \aap, 538, A47

\bibitem[{{Nota} {et~al}\mbox{.}(1995){Nota}, {Livio}, {Clampin}, \&
  {Schulte-Ladbeck}}]{nota95}
{Nota} A., {Livio} M., {Clampin} M., {Schulte-Ladbeck} R., 1995, \apj, 448, 788

\bibitem[{{Pamyatnykh}(1999)}]{pam_1999}
{Pamyatnykh} A.~A., 1999, \actaa, 49, 119

\bibitem[{{Pojmanski}(1997)}]{ASAS}
{Pojmanski} G., 1997, \actaa, 47, 467

\bibitem[{{Ritchie} {et~al}\mbox{.}(2009){Ritchie}, {Clark}, {Negueruela}, \&
  {Crowther}}]{ritchie}
{Ritchie} B.~W., {Clark} J.~S., {Negueruela} I., {Crowther} P.~A., 2009, \aap,
  507, 1585

\bibitem[{{Saio} {et~al}\mbox{.}(2013){Saio}, {Georgy}, \&
  {Meynet}}]{saio_2013}
{Saio} H., {Georgy} C., {Meynet} G., 2013, \mnras, 433, 1246

\bibitem[{{Saio} {et~al}\mbox{.}(2006){Saio}, {Kuschnig}, {Gautschy},
  {Cameron}, {Walker}, {Matthews}, {Guenther}, {Moffat}, {Rucinski},
  {Sasselov}, \& {Weiss}}]{saio_2006}
{Saio} H. {et~al.}, 2006, \apj, 650, 1111

\bibitem[{{Sana} {et~al}\mbox{.}(2012){Sana}, {de Mink}, {de Koter}, {Langer},
  {Evans}, {Gieles}, {Gosset}, {Izzard}, {Le Bouquin}, \&
  {Schneider}}]{sana_science}
{Sana} H. {et~al.}, 2012, Science, 337, 444

\bibitem[{{Sana} {et~al}\mbox{.}(2009){Sana}, {Gosset}, \&
  {Evans}}]{sana_binarityII}
{Sana} H., {Gosset} E., {Evans} C.~J., 2009, \mnras, 400, 1479

\bibitem[{Sana {et~al}\mbox{.}(2006)Sana, Gosset, \& Rauw}]{sana06stars}
Sana H., Gosset E., Rauw G., 2006, \mnras, 371, 67

\bibitem[{{Sana} {et~al}\mbox{.}(2011){Sana}, {James}, \&
  {Gosset}}]{sana_binarityIII}
{Sana} H., {James} G., {Gosset} E., 2011, \mnras, 416, 817

\bibitem[{{Sher}(1965)}]{sher}
{Sher} D., 1965, \mnras, 129, 237

\bibitem[{{Sim{\'o}n-D{\'{\i}}az} \& {Herrero}(2014)}]{sergio_2013}
{Sim{\'o}n-D{\'{\i}}az} S., {Herrero} A., 2014, \aap, 562, A135

\bibitem[{{Sim{\'o}n-D{\'{\i}}az} {et~al}\mbox{.}(2010){Sim{\'o}n-D{\'{\i}}az},
  {Herrero}, {Uytterhoeven}, {Castro}, {Aerts}, \& {Puls}}]{sergio_lpvs}
{Sim{\'o}n-D{\'{\i}}az} S., {Herrero} A., {Uytterhoeven} K., {Castro} N.,
  {Aerts} C., {Puls} J., 2010, \apjl, 720, L174

\bibitem[{{Smartt} {et~al}\mbox{.}(2002){Smartt}, {Lennon}, {Kudritzki},
  {Rosales}, {Ryans}, \& {Wright}}]{smartt_2002}
{Smartt} S.~J., {Lennon} D.~J., {Kudritzki} R.~P., {Rosales} F., {Ryans}
  R.~S.~I., {Wright} N., 2002, \aap, 391, 979

\bibitem[{{Smith}(2007)}]{smith_HD168625}
{Smith} N., 2007, \aj, 133, 1034

\bibitem[{{Smith} {et~al}\mbox{.}(2013){Smith}, {Arnett}, {Bally}, {Ginsburg},
  \& {Filippenko}}]{sbw1_2013}
{Smith} N., {Arnett} W.~D., {Bally} J., {Ginsburg} A., {Filippenko} A.~V.,
  2013, \mnras, 429, 1324

\bibitem[{{Smith} {et~al}\mbox{.}(2007){Smith}, {Bally}, \&
  {Walawender}}]{darkness}
{Smith} N., {Bally} J., {Walawender} J., 2007, \aj, 134, 846

\bibitem[{{Smith} \& {Townsend}(2007)}]{smith_townsend}
{Smith} N., {Townsend} R.~H.~D., 2007, \apj, 666, 967

\bibitem[{{Walborn} \& {Fitzpatrick}(2000)}]{nolan_2000}
{Walborn} N.~R., {Fitzpatrick} E.~L., 2000, \pasp, 112, 50

\bibitem[{{Walborn} {et~al}\mbox{.}(1989){Walborn}, {Prevot}, {Prevot},
  {Wamsteker}, {Gonzalez}, {Gilmozzi}, \& {Fitzpatrick}}]{nolan_87a}
{Walborn} N.~R., {Prevot} M.~L., {Prevot} L., {Wamsteker} W., {Gonzalez} R.,
  {Gilmozzi} R., {Fitzpatrick} E.~L., 1989, \aap, 219, 229

\bibitem[{{Wolfe} {et~al}\mbox{.}(1967){Wolfe}, {Horak}, \&
  {Storer}}]{wolfe_orbits}
{Wolfe}, Jr. R.~H., {Horak} H.~G., {Storer} N.~W., 1967, {The machine
  computation of spectroscopic binary elements}, Gordon \& Breach, p. 251

\bibitem[{{Zima}(2008)}]{zima_FAMIAS}
{Zima} W., 2008, Communications in Asteroseismology, 157, 387

\end{thebibliography}

\begin{table*}
\renewcommand\thetable{A1} 
\begin{small}
\caption{Radial velocities (RVs) for the spectral lines used to characterise the three target objects. The RVs (in kms$^{-1}$) are relative to the mean RV of all the measurements of that object across all epochs. Dashes indicate lines where RVs were not estimated (generally due to poor signal-to-noise). Where back-to-back frames were co-added, the date given is the mid-point of the combined frames. The full version of this table is available online.
\label{measurements} } 
\begin{center} 
\begin{tabular}{| lr |r |r |r |r |r |r |r |r |r |r |r |r |r |r |r |r |r |}
\hline\hline 
Date  &          4026  &          4144  &          4553  &          4568  & 
        4575  &          4591  &          4607  &          4631  &          5011
 &          5667  &          5676  &          8598  &          8665  &   Mean
 &   Stddev  \\
{\tiny (HJD - 240000)} & He\,I  & He\,I & Si\,III & Si\,III & Si\,III & O\,II & N\,II & N\,II & O\,III & N\,II & C\,I \ & H \ \  & H \ \  \\
\hline
\vspace{0.15cm}
{\it \hspace{-0.3cm} Sher25} \\
54909.05  &   -   &   -   &     -4.32  &     -6.46  &     -9.12  &     -8.24  & 
   -8.27  &     -4.71  &     -8.97  &     -3.30  &     -7.20  &     -5.89  & 
   -4.04  &     -6.41  &      2.10  \\
54909.12  &   -   &   -   &     -7.32  &     -6.35  &     -7.08  &     -8.93  & 
   -8.21  &     -3.39  &     -7.66  &     -3.45  &     -6.66  &     -6.00  & 
   -4.98  &     -6.37  &      1.81  \\
54909.22  &   -   &   -   &     -6.74  &     -7.49  &    -10.84  &     -5.31  & 
   -4.76  &     -4.69  &     -7.45  &     -4.33  &     -8.19  &     -8.47  & 
   -5.08  &     -6.67  &      2.04  \\
... &      &     &      &      &      &      &      &     &     &     &     &     &    &     &       \\
\hline
\end{tabular} 
\end{center} 
\end{small}
\end{table*}

\end{document}